\input{aipcheck}

\documentclass[
    ,final            % use final for the camera ready runs
    ,numberedheadings % uncomment this option for numbered sections
  ]
  {aipproc}

\layoutstyle{8x11single}

\begin{document}

\title{WARM INFLATION IN PRESENCE OF MAGNETIC FIELDS}


\author{Gabriella Piccinelli}{
  address={Centro Tecnológico, FES Aragón, Universidad Nacional Autónoma de México, Avenida Rancho Seco S/N, Bosques de Aragón, Nezahualcóyotl, Estado de México 57130, Mexico}
}

\author{\'Angel S\'anchez}{
  address={Department of Physics, University of Texas at El Paso, El Paso, Texas 79968, USA}
}

\author{Alejandro Ayala}{
  address={Instituto de Ciencias Nucleares, Universidad Nacional Autónoma de México, Apartado Postal 70-543, México Distrito Federal 04510, Mexico}
}

\author{Ana Julia Mizher}{
  address={Instituto de Ciencias Nucleares, Universidad Nacional Autónoma de México, Apartado Postal 70-543, México Distrito Federal 04510, Mexico}
}

\begin{abstract}
We present preliminary results on the possible effects that primordial magnetic fields can have for a warm inflation scenario, based on global supersymmetry, with a new-inflation-type potential. This work is motivated by two considerations: first, magnetic fields seem to be present in the universe on all scales which rises de possibility that they could also permeate the early universe; second, the recent emergence of inflationary models where the inflaton is not assumed to be isolated but instead it is taken as an interacting field, even during the inflationary expansion.

The effects of magnetic fields are included resorting to Schwinger's proper time method.
\end{abstract}

\maketitle

%%%%%%%%%%%%%%%%%%%%%%%%%%%%%%%%%%%%%%%%%%%%
%% MAINMATTER
%%%%%%%%%%%%%%%%%%%%%%%%%%%%%%%%%%%%%%%%%%%%

\section{INTRODUCTION}

\paragraph{PRIMORDIAL MAGNETIC FIELDS} Magnetic fields seem to be ubiquitous in the universe, though their origin is at present unknown. In view of the accumulating observational evidence for their presence on all scales (for comprehensive reviews see \cite{Kronberg94}), up to galaxy clusters \cite{Clarke02} and superclusters \cite{Kim89}, the idea of a truly primordial origin of cosmic magnetism gains strength. If this is the case, two important issues have to be addressed. On one side, successful mechanisms for the cosmological generation, preservation and amplification of magnetic fields have to be found, on the other, a complete cosmological model that includes the effects of magnetic fields has yet to be developed.

A series of mechanisms have been proposed for the early generation of magnetic fields (\cite{Turner88}, \cite{Widrow}, \cite{Dolgov10}, and \cite{Kandus91} for a review) but none of them is problem free. It is difficult to obtain both the required scale and amplitude to match the presently observed fields. Nonetheless, even if these fields do not survive up to the present epoch, there are good chances that they were present during the early stages of the universe evolution, in particular, during the inflationary epoch where large-scale cosmological magnetic fields can be created through the same mechanism that generates density fluctuations: quantum fluctuations in the Maxwell field are excited inside the horizon and are expected to freeze-out as classical electromagnetic waves once they cross the Hubble radius. These initially static electric and magnetic fields can subsequently lead to current supported magnetic fields, once the excited modes reenter the horizon.  Nevertheless, magnetic fluctuations that have survived a period of de Sitter expansion are typically too weak to match the present observations, since magnetic fields decay adiabatically with the universe expansion. Thus, the conformal invariance of electromagnetism must be broken to avoid this huge suppression and this can be achieved in models where the electromagnetic field is coupled to gravity \cite{Turner88}. Another possibility is that the curvature of the background space can modifiy this adiabatic decay law \cite{Tsagas05}. Also, non-Abelian gauge theories may have a ferromagnet-like vacuum (Savvidy vacuum), with a non-zero magnetic field, even at high temperatures \cite{Savvidy77}. The formation of this non-trivial vacuum state at GUT scales can give rise to a Maxwell magnetic field imprinted on the comoving plasma.

The influence of magnetic fields becomes important whenever the universe presents suitable conditions for charge separation and motion during its thermal evolution. Their effect has been widely studied on the cosmic background radiation (see e.g. \cite{Yamazaki69} and references there in); on nucleosyhtesis processes (a detailed review can be found in \cite{Grasso01}); on cosmological phase transitions \cite{Giovanini98}, \cite{Sanchez07} and, in particular, on the baryogenesis process (\cite{Piccinelli04} for a review, see also \cite{Comelli99}); and on structure formation \cite{Widrow}. Nevertheless, the implications on the fate of the cosmological inflation potential have not yet, to our knowledge, been considered. It is the aim of this work to contribute to the building of the magnetic cosmological model introducing magnetic fields in this very early epoch.

\paragraph{WARM INFLATION} The early models of inflation -dubbed super-cooled models (see e.g. \cite{Olive99} for a review)- assumed a small interaction of the inflaton with all other fields until the reheating process, at the end of inflation. With the proposal of warm inflation \cite{Berera95} the picture changed: the inflaton field is supposed to be interacting with other fields, in a continuous and therefore more natural way, during both the inflationary expansion and the reheating. In this model thermal equilibrium is maintained during the inflationary expansion, with no need for a tiny coupling constant. It does require a dissipative component $\Gamma$ of a valuable size as compared to the universe expansion rate, as opposed to the standard inflationary process, where the damping term is provided only by the universe's expansion. In this way, the equation of motion for the inflaton $\phi$ becomes:
\begin{equation}
\ddot\phi+(3H+\Gamma)\dot\phi+V_{T,\phi}=0,\label{wip}
\end{equation}
where $H$ is the Hubble parameter, with $\Gamma>3H$, and $V_{T, \phi}$ is the derivative with respect to $\phi$ of the inflaton effective potential, taken to be the finite temperature one-loop Coleman-Weinberg potential.

Since the inflaton interacts with particles during the inflationary process, it seems also natural to consider its interaction with all kind of fields present at that time, in particular magnetic fields. Consequently, thermal and magnetic corrections to the inflaton potential have to be considered. Thermal effects where worked in \cite{Hall04}. Here we address the problem of the magnetic field contributions.

\section{Thermal and Magnetic effects}

\paragraph{THE MODEL} We will closely follow the work done by Hall and Moss [18], where they compute thermal corrections to a successful version of the warm inflation model and show that the flatness of the inflationary potential is not spoiled by finite temperature effects. They work with a particle model in the context of global supersymmetry, when the dissipative effects of particle production are taken into account.

They start from a new-inflation-type potential and work with three superfields: $\Phi$, whose scalar component $\varphi$ is identified with the inflaton ($\phi = \sqrt 2 Re \varphi$), $X$, whose scalar component $\chi$ is coupled to the inflaton and becomes very heavy and $Y$, with a vanishing coupling, who gives rise to the light sector. In a two stage reheating process, $\phi \rightarrow \chi \rightarrow \tilde y \tilde y$, with $\tilde y$ a light fermion,  the radiative corrections to the inflaton potential are shown to be small due to fermion-boson cancellation and to the fact that thermal contribution to the inflaton mass from heavy sector loops are Boltzmann suppressed. Further assumptions are that there is a soft SUSY breaking in the heavy sector and that light radiation thermalizes.

We are interested in the full set of interactions that involve the inflaton and the $\chi$ field, that can be read in the following sectors (scalar and fermionic) of the Lagrangian:
\begin{eqnarray}
{\cal L}_s & = &g^2 |\Lambda^2-|\chi|^2|^2 +4g^2 |\varphi|^2|\chi|^2
+4h^2|y|^2|\chi|^2+h^2|y|^4
+2gh(y^2\varphi^\dagger\chi^\dagger+y^{\dagger2}\varphi\chi) \nonumber \\
{\cal L}_f &= &g(\varphi \overline \psi_{\chi} P_L\psi_{\chi}+
\varphi^\dagger \overline \psi_{\chi}P_R \psi_{\chi})
+h(\chi \overline\psi_{y}
P_L\psi_{y} + \chi^\dagger \overline\psi_{y} P_R \psi_{y})\nonumber\\
&&+2g(\chi\overline\psi_{\chi}P_L \psi_{\varphi} +
\chi^\dagger\overline\psi_{\chi} P_R \psi_{\varphi}) 
+2h(y\overline\psi_{y} P_L\psi_{\chi} + y^\dagger\overline\psi_{y}
P_R\psi_{\chi}),
\label{lagrangian}
\end{eqnarray}
 where $y$ is the scalar field component of the chiral  superfield $Y$,  $\Psi_i$ denote the fermionic fields of the different sectors, $P_L = 1-P_R= (1+ \gamma_5)/2$, $g$ and $h$ are coupling constants and $\Lambda$ is a mass scale. Normalising the density perturbation amplitude to the cosmic microwave background leads to coupling constants $g$ and $h$ around 0.1 and a mass scale of up to $10^{11} GeV$.

\begin{figure}
   \label{fig1}
  \includegraphics[height=.2\textheight]{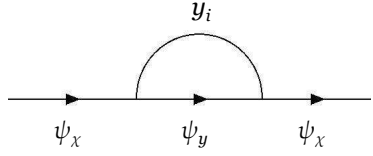}
  \caption{Feynman diagram for the heavy fermionic sector self-energy.}
\end{figure}

\begin{figure}
   \label{fig2}
  \includegraphics[height=.4\textheight]{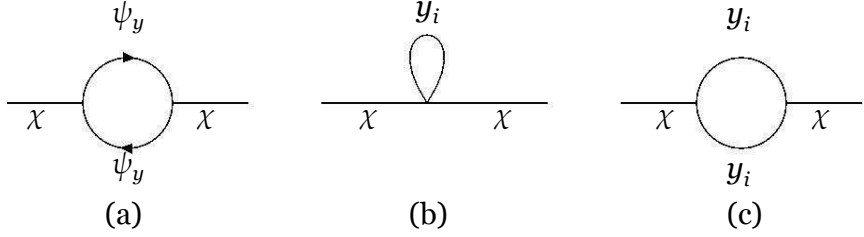}
  \caption{Feynman diagrams for the heavy bosonic sector self-energy.}
\end{figure}

Thermal and magnetic corrections to the inflaton potential will appear as a result of the self-energies of the $\chi$ and $\Psi_{\chi}$ fields. Since the light sector thermalises and the heavy one does not, a series of simplifications can be adopted: inside the self-energy loops, we can deal with the light sector through a Hard Thermal Loop (HTL) approximation (see e.g.\cite{Bellac96}), while the heavy sector is suppressed; outside the loop, we have $\chi$ and $\Psi_{\chi}$ fields, with $T \ll m_{\chi}, m_{\Psi_{\chi}}$. In such a way, the only Feynman diagrams that we need for computing the thermal and magnetic corrections to the bosonic and fermionic masses are the ones depicted in Fig.~\eqref{fig1} and Fig.~\eqref{fig2}.

%%%%%%%%%%%%%%%%%%%%%%%%%%%%%%%%%%%%%%%%%%%%
%% Figuras:
%%

%%
%% 
%%%%%%%%%%%%%%%%%%%%%%%%%%%%%%%%%%%%%%%%%%%%

Working in imaginary time formalism and adopting the notation that 4-momenta are written in upper case and 3-momenta in lower case, the fermionic self-energy, corresponding to Fig.~\eqref{fig1} is \cite{Hall04}, \cite{Bellac96}:
\begin{equation}
\label{3momenta}
\Sigma(P) = -4 h^2\,T\,\sum_n\int \frac{d^3k}{(2 \pi)^3}
(K\!\!\!\!/ - P\!\!\!\!/) \Delta(K) \widetilde{\Delta}(P-K),
\end{equation}
where $\Delta(K)\approx K^{-2}$, $k^0=2n\pi T$ for bosons and $k^0=(2n+1)\pi T$
for fermions (denoted by a tilde), which, in the infrared limit ($p_0=0$, $p \rightarrow 0$), leads to:
\begin{equation}
m_f^2 \equiv \Sigma \approx \frac{h^2 T^2}{2}.
\end{equation}

The bosonic self-energy is composed by the contributions of the three diagrams in Fig.~\eqref{fig2}
\begin{equation}
\Pi(P)_a = h^2\,T\,\sum_n\int \frac{d^3k}{(2 \pi)^3}
Tr\left[K\!\!\!\!/(K\!\!\!\!/ - P\!\!\! \!/)\right] 
\widetilde{\Delta}(K) \widetilde{\Delta}(K-P).
\end{equation}
In the HTL limit, this leads to:
\begin{equation}
\label{HTLeq}
\Pi(P)_a= -4 h^2 \,T\,\sum_n\int \frac{d^3k}{(2 \pi)^3}K^2
\widetilde{\Delta}(K) \widetilde{\Delta}(K-P) \approx {1 \over 6}h^2 T^2.
\end{equation}

Similarly,
\begin{equation}
\label{finitetempself}
\Pi(P)_b = 4 h^2\,T\,\sum_n\int \frac{d^3k}{(2 \pi)^3}
{\Delta}(K) 
\approx \frac13h^2T^2.
\end{equation}

\begin{eqnarray}
\Pi(P)_c &=& 4g^2 h^2\phi^2\,T\,\sum_n\int \frac{d^3k}{(2 \pi)^3}{\Delta}(K) {\Delta}(K-P) \nonumber \\
 &\approx& {1\over 2\pi^2}g^2 h^2\phi^2\log{T^2\over p^2}.
\end{eqnarray}
This contribution of the self-energy is associated to a finite temperature correction to the coupling constant $g(T)$.

Together, diagrams (a) and (b) define the thermal contribution to the bosonic mass:

\begin{equation}
m^2_b \equiv \Pi_a + \Pi_b \approx \frac{h^2 T^2}{2}.
\end{equation}

\paragraph{CONTRIBUTION OF MAGNETIC FIELDS} We work with a constant magnetic field of strength $B$ along the $z$ axis and with the assumption that the hierarchy of scales $eB<<m_i^2<<T^2$ is obeyed, where $m_i$ is the mass of the fields inside the loop.

To include the effect of an external magnetic field, we work with Schwinger proper time method \cite{Schwinger51}, where the momentum dependent propagators for charged scalars and fermions coupled to the external field take, respectively, the form:
\begin{eqnarray}
\label{scalpropmom}
i D_{B} (k) &=& \int_0^\infty \frac{ds}{\cos{eBs}} \nonumber \\
&& \times \exp\left\{ i s (k_{||}^2-k_{\bot}^2 \frac{\tan{eBs}}{eBs}-m^2_b +i
\epsilon)\right\},
\end{eqnarray}
\begin{eqnarray}
i S_{B} (k)&=&\int_0^\infty \frac{ds}{\cos{eBs}} \nonumber \\
& & \times \exp\left\{ i s (k_{||}^2-k_{\bot}^2 \frac{\tan{eBs}}{ eBs}-m^2_{f}
+i \epsilon)\right\} \nonumber \\
& & \times \left[ (m_f+{\not \! k}_{||})e^{i eB s \sigma_3} -\frac{{\not \!
k_\bot}}{\cos{eB s}}\right], \label{ferpropmom}
\end{eqnarray}
with the notation $k_{||}^2=k_0^2-k_3^2$, $k_\perp^2=k_1^2+k_2^2$, and $\sigma^3=i\gamma^1\gamma^2=-\gamma^5{\not \! u}{\not \! b}$, where $\not \! u$ and $\not \! b$ are four-vectors describing the plasma rest frame and the direction of the magnetic field, respectively.

It has been shown that, by deforming the contour of integration,
Eqs.~(\ref{scalpropmom}) and~(\ref{ferpropmom}) can be written
as \cite{Ayala05}
\begin{eqnarray}
iD_{B}(k)=2i\sum_{l=0}^{\infty}\frac{(-1)^lL_l(\frac{2k_\perp^2}{e B})
{\mathrm e}^{-\frac{k^2_\perp}{e B}}}{k^2_{||}-(2l+1)e B-m^2_b+i\epsilon},
\label{scalpropsum}
\end{eqnarray}
\begin{eqnarray}
iS_{B}(k)= i \sum^\infty_{l=0} \frac{d_l(\frac{k_\perp^2}{e B})D +
d^{\prime}_l(\frac{k_\perp^2}{e B}) \bar D}{k^2_{||}-2 l e B-m_f^2 +
i\epsilon} + \frac{{\not \! k_{\bot}}}{k^2_\perp}, \label{ferpropsum}
\end{eqnarray}
where $d_l(\alpha)\equiv (-1)^n e^{-\alpha} L^{-1}_l(2\alpha)$,
$d^{\prime}_n=\partial d_n/\partial \alpha$,
\begin{eqnarray}
D &=& (m_f+{\not \! k_{||}})+ {\not \! k_{\perp}} \frac{m_f^2-k^2_{||}}{ {
k^2_{\perp}}}, \nonumber \\
\bar D &=& \gamma_5 {\not \! u}{\not \! b}(m_f + {\not \! k_{||}}),
\label{DDe}
\end{eqnarray}
and $L_l$, $L_l^m$ are Laguerre and Associated Laguerre polynomials,
respectively.

We can thus perform a weak field expansion in Eqs.~(\ref{scalpropsum}) and (\ref{ferpropsum}), which allows us to carry out the summation over Landau levels to write the scalar and fermion propagators as power series in $eB$, that up to order $(eB)^2$ read as
\begin{eqnarray}
\label{scalpropweak}
D_{B}(k)=\frac{1}{k^2-m^2_b}\left( 1-\frac{(eB)^2}{(k^2-m^2_b)^2}- \frac{2(eB)^2
k_\perp^2}{(k^2-m^2_b)^3}\right),
\end{eqnarray}

\begin{equation}
{S_{B}(k)} = \frac{{\not \! k}+m_f}{{\not \! k}^2-m_f^2}+
\frac{\gamma_5 {\not \! u}{\not \! b}(k_{||}+m_f)(eB)}{(k^2-m_f^2)^3}
-\frac{2(eB)^2 k_\perp^2}{(k^2-m_f^2)^4} (m_f+{\not \! k_{||}}+{\not \!
k_\perp}\frac{m_f^2-k_{||}^2}{k_\perp^2}).
\label{ferpropweak}
\end{equation}

Using these propagators in the self-energies (Eqs.~\eqref{3momenta},~\eqref{HTLeq} and~\eqref{finitetempself}), we obtain the leading order corrections, from thermal and magnetic effects, to the bosonic and fermionic masses of the heavy sector
\begin{equation} 
m^2_b (T,B) \approx \frac{h^2 T^2}{2} \left( 1 - \frac{2 m_y}{\pi T} - \frac{m^2_y}{2 \pi^2 T^2} \left[\ln \left( \frac{m^2_y}{(4 \pi T)^2} \right)+2 \gamma_{_E} - 1 \right] - {1 \over 12 \pi} \frac{(eB)^2}{m^3_y T} \right)
\end{equation}
\begin{equation} 
m^2_f (T,B) \approx \frac{h^2 T^2}{2} \left( 1 - {1 \over 3} \frac{r (eB)}{\pi m_y T} + {11 \over 12 \pi} \frac{(eB)^2}{m^3_y T} \right),
\end{equation}
where $r= \pm 1$ represents the two posible spin orientations of the fermion with respect to the magnetic field.

These self-energies will now correct the boson and fermion propagators
\begin{equation}
i S^{-1} = P\!\!\!\!/ - m^2_{\Psi_\chi},
\label{propagador1}
\end{equation}
\begin{equation}
G^{-1} = P^2 + m^2_{\chi},
\label{propagador2}
\end{equation}
which, in turn, will modify the effective potential of the inflaton
\begin{equation}
V_\chi =\int \frac{d^4P}{(2\pi)^4} \ln\\\\\ \det(G^{-1})
-\int \frac{d^4P}{(2\pi)^4} \ln\\\\\ \det(S^{-1}S^{*-1})^{-1/2}.
\label{potencial1}
\end{equation}
This potential has to be computed taking into account a soft SUSY breaking that will introduce an extra term ($M_s$) in the boson mass
\begin{eqnarray}
m^2_\chi&=&2g^2\phi^2+ m_b^2(T,B) + M_s^2, \nonumber \\
m^2_{\Psi_\chi}&=&2g^2\phi^2 + m_f^2(T,B).
\label{masas}
\end{eqnarray}

Finally, the complete inflaton potential will be
\begin{equation} 
V(\phi,T, B) = -\frac{\pi^2}{90} g_{*}T^4 + V_\chi(\phi, T, B).
\label{potencial2}
\end{equation}
The function $V_\chi$ can be explicitly computed and the final results will be reported elsewhere \cite{PiccinelliP}.

In the weak field approximation, the effect of magnetic fields on the features of the effective potential, such as its flatness and the location of the minimum, is expected to be small.

\section{CONCLUSIONS AND PERSPECTIVES}
 
In the context of a warm inflation model embedded in global supersymmetry, we have introduced magnetic fields in thermal propagators of charged fermions and scalars and computed the leading corrections to the self-energies of these particles. Our aim is to introduce these contributions in the inflaton effective potential and analyze if it fulfills the slow-roll conditions, thus establishing the range of the magnetic field intensity that does not spoil the inflaton potential flatness. This allows us to explore if warm inflation can impose some bounds on primordial magnetic fields and vice versa. As a second step, the effect on the density fluctuations spectrum will be studied.

%%%%%%%%%%%%%%%%%%%%%%%%%%%%%%%%%%%%%%%%%%%%%%%%
%% BACKMATTER
%%%%%%%%%%%%%%%%%%%%%%%%%%%%%%%%%%%%%%%%%%%%%%%%

\begin{theacknowledgments}

G.P. aknowledges support received by DGAPA-UNAM under PAPIIT grant IN117111.
 
\end{theacknowledgments}



\bibliographystyle{aipproc}   % if natbib is available

\bibliography{sample}

\IfFileExists{\jobname.bbl}{}
 {\typeout{}
  \typeout{******************************************}
  \typeout{** Please run "bibtex \jobname" to optain}
  \typeout{** the bibliography and then re-run LaTeX}
  \typeout{** twice to fix the references!}
  \typeout{******************************************}
  \typeout{}
 }

%%\end{document}

%%%%%%%%%%%%%%%%%%%%%%%%%%%%%%%%%%%%%%%%%%%
%% The following lines show an example how to produce a bibliography
%% without the help of the BibTeX program. This could be used instead
%% of the above.
%%%%%%%%%%%%%%%%%%%%%%%%%%%%%%%%%%%%%%%%%%%

\end{document}